%% file: susyewpo_hepph.tex
\documentclass[12pt]{article}
\usepackage{epsfig,cite,amsmath}

\makeatletter

\def\paragraph{\@startsection{paragraph}{4}{\z@}{+2.00ex plus
 +1ex minus +.2ex}{1.5ex plus .2ex}{\it\normalsize}}

\def\section{\@startsection {section}{1}{\z@}{+3.0ex plus +1ex minus
  +.2ex}{2.3ex plus .2ex}{\normalsize\bf\boldmath}}
\def\subsection{\@startsection{subsection}{2}{\z@}{+2.5ex plus +1ex
minus +.2ex}{1.5ex plus .2ex}{\normalsize\bf\boldmath}}
\def\subsubsection{\@startsection{subsubsection}{3}{\z@}{+3.25ex plus
 +1ex minus +.2ex}{1.5ex plus .2ex}{\normalsize\bf\boldmath}}

\expandafter\ifx\csname mathrm\endcsname\relax\def\mathrm#1{{\rm #1}}\fi

\newcounter{saveeqn}

\@addtoreset{equation}{section}

\input paperdef

\begin{document}

\thispagestyle{empty}
\setcounter{page}{0}
\def\thefootnote{\fnsymbol{footnote}}

\begin{flushright}
DCPT/03/84\\
IPPP/03/42\\
LMU 14/03\\
hep-ph/0307177\\
\date{\today}
\end{flushright}

\vspace{1cm}

\begin{center}

{\large\sc {\bf The MSSM in the Light of Precision Data}}
\footnote{to appear in the proceedings of the workshop
``Electroweak precision data and the Higgs mass'', 
\mbox{}~~~~~~DESY Zeuthen, February 2003
}

\vspace{1cm}

{\sc 
S.~Heinemeyer$^{1}$%
\footnote{email: Sven.Heinemeyer@physik.uni-muenchen.de}%
~and G.~Weiglein$^{2}$%
\footnote{email: Georg.Weiglein@durham.ac.uk}
}

\vspace*{1cm}

{\sl
$^1$Institut f\"ur theoretische Elementarteilchenphysik,
LMU M\"unchen, Theresienstr.\ 37, D-80333 M\"unchen, Germany

\vspace*{0.4cm}

$^2$Institute for Particle Physics Phenomenology, University of Durham,\\
Durham DH1~3LE, UK

}

\end{center}

\vspace*{1cm}

\begin{abstract}

The potential of present and anticipated future electroweak precision data, 
including the Higgs boson and top quark masses, for testing quantum effects 
of the electroweak theory is investigated in the context of the Minimal 
Supersymmetric Standard Model (MSSM). The present status of the
theoretical predictions is analyzed. The impact of the parametric
uncertainties from the experimental errors of the input parameters is
studied, and an estimate for the remaining uncertainties from unknown
higher-order corrections is given both in the Standard Model (SM) and
the MSSM. Examples of electroweak precision tests in the mSUGRA
scenario and the unconstrained MSSM are 
analyzed, and the status of the global fit to all data is discussed.
 
\end{abstract}

\def\thefootnote{\arabic{footnote}}
\setcounter{footnote}{0}

\newpage



\vspace*{1cm}
\begin{center}

{\Large \bf  The MSSM in the Light of Precision Data}

\vspace*{1cm}

{\sc S.~Heinemeyer$^{1}$ and G.~Weiglein$^{2}$}

\vspace*{.5cm}

{\normalsize \it
$^1$Institut f\"ur theoretische Elementarteilchenphysik,
LMU M\"unchen, Theresienstr.\ 37, D--80333 M\"unchen, Germany\\[.2em] 
$^2$Institute for Particle Physics Phenomenology, University of Durham,\\
Durham DH1~3LE, UK
}
\par
\end{center}
\vskip 1cm
\begin{center}
\bf Abstract
\end{center} 
{\it
The potential of present and anticipated future electroweak precision data, 
including the Higgs boson and top quark masses, for testing quantum effects 
of the electroweak theory is investigated in the context of the Minimal 
Supersymmetric Standard Model (MSSM). The present status of the
theoretical predictions is analyzed. The impact of the parametric
uncertainties from the experimental errors of the input parameters is
studied, and an estimate for the remaining uncertainties from unknown
higher-order corrections is given both in the Standard Model (SM) and
the MSSM. Examples of electroweak precision tests in the mSUGRA
scenario and the unconstrained MSSM are 
analyzed, and the status of the global fit to all data is discussed.
}
\par
\vskip 1cm


\section{Introduction}

Theories based on Supersymmetry (SUSY) \cite{susy} are widely
considered as the theoretically most appealing extension of the
Standard Model (SM). They are consistent with the approximate
unification of the gauge coupling constants at the GUT scale and
provide a way to cancel the quadratic divergences in the Higgs sector
hence stabilizing the huge hierarchy between the GUT and the Fermi
scales. Furthermore, in SUSY theories the breaking of the electroweak
symmetry is naturally induced at the Fermi scale, and the lightest
supersymmetric particle can be neutral, weakly interacting and
absolutely stable, providing therefore a natural solution for the dark
matter problem. 
SUSY predicts the existence of scalar partners $\tilde{f}_L, 
\tilde{f}_R$ to each SM chiral fermion, and spin--1/2 partners to the 
gauge bosons and to the scalar Higgs bosons. So far, the direct search for
SUSY particles has not been successful. 
One can only set lower bounds of ${\cal O}(100)$~GeV on 
their masses~\cite{pdg}. 

An alternative way to probe SUSY is via the virtual effects of the 
additional particles to precision observables. 
This requires a very high precision 
of the experimental results as well as of the theoretical predictions.
The most relevant electroweak precision observables (EWPO) in this
context are the $W$~boson mass, $\MW$, the effective leptonic weak mixing
angle, $\sweff$, and the mass of the lightest $\cp$-even MSSM Higgs boson,
$\mh$. Contrary to the SM case, where the mass of the Higgs boson is a
free parameter, within the MSSM the quartic
couplings of the Higgs potential are fixed in terms of the gauge
couplings as a consequence of SUSY~\cite{hhg}. Thus, at the
tree-level, the Higgs sector  
is determined by just two independent parameters besides the SM
electroweak gauge couplings $g$ and $g'$, conventionally chosen as
$\tb = v_2/v_1$, the ratio of the vacuum expectation values of the two
Higgs doublets, and $\MA$, the mass of the $\cp$-odd $A$~boson. 
As a consequence, the mass of the lightest $\cp$-even MSSM Higgs boson can 
be predicted in terms of the other model parameters. 

An upper bound of $\mh \lsim 135 \gev$~\cite{mhiggslong,mhiggsAEC} can
be established, taking into 
account all existing higher-order corrections (for $\mt = 175 \gev$
and a common soft SUSY-breaking scale of $\msusy = 1 \tev$). The
prospective accuracy of the measurement of the Higgs-boson mass at the
LHC of about $200 \mev$~\cite{lhc} or at an \epem\ linear collider (LC)
of even $50 \mev$~\cite{teslatdr,nlc,jlc} will promote $\mh$ to a
precision observable. Owing to the sensitive dependence of $\mh$ on 
especially the scalar top sector, the measured value of $\mh$ will allow
to set stringent constraints on the parameters in this sector.

In the unconstrained MSSM no specific assumptions are made about the 
underlying SUSY-breaking mechanism, and a parameterization of all 
possible SUSY-breaking terms is used. This gives rise to the huge number of 
more than 100 new parameters in addition to the SM ones, 
which in principle can be chosen
independently of each other. A phenomenological analysis of this model
in full generality would clearly be very involved, and one usually
restricts to certain benchmark scenarios, see e.g.\
\citeres{benchmark,benchmark2,sps}.
On the other hand, models in which all the low-energy parameters are
determined in terms of a few parameters at the Grand Unification
scale (or another high-energy scale), 
employing a specific soft SUSY-breaking scenario, are
much more predictive. The most prominent scenarios in the literature
are minimal Supergravity (mSUGRA)~\cite{susy}, minimal Gauge Mediated
SUSY Breaking (mGMSB)~\cite{gmsbrev} and minimal Anomaly Mediated SUSY
Breaking (mAMSB)~\cite{amsb1,amsb2,amsb3}.  
Analyses comparing the Higgs sector in these scenarios and discussing
implications for searches at present and future colliders can be found
in \citeres{asbs,asbs2}. 

Examples for the current experimental 
status of EWPO are given in \refta{tab:ewpotoday},
including their relative experimental precision. The quantities in the
first three lines, $\MZ$, $\gf$, and $\mt$, are usually employed as
input parameters for the theoretical predictions. The observables $\MW$,
$\sweff$, $\Ga_Z$, on the other hand, are the three most prominent
observables for testing the electroweak theory by comparing the
experimental results with the theory predictions. Comparing the typical
size of electroweak quantum effects, which is at the per cent level,
with the relative accuracies in \refta{tab:ewpotoday}, which are at the
per mille level, clearly shows the sensitivity of the electroweak precision
data to loop effects.
%
\begin{table}[htb]
\renewcommand{\arraystretch}{1.5}
\BC
\begin{tabular}{|c||c|c|c|}
\cline{2-4} \multicolumn{1}{c||}{}
& central value & absolute error & relative error \\
\hline\hline
$\MZ$ [GeV] & 91.1875 & $\pm 0.0021$ & $\pm 0.002\%$ \\ \hline
$\gf$ [GeV$^{-2}$] & $1.16637 \times 10^{-5}$ & $\pm 0.00001 \times 10^{-5}$
                   & $\pm 0.0009\%$ \\ \hline
$\mt$ [GeV] & 174.3 & $\pm 5.1$ & $\pm 2.9 \%$ \\ \hline
$\MW$ [GeV] & 80.426 & $\pm 0.034$ & $\pm 0.04\%$ \\ \hline
$\sweff$ & 0.23148 & $\pm 0.00017$ & $\pm 0.07\%$ \\ \hline
$\Ga_Z$ [GeV] & 2.4952 & $\pm 0.0023$ & $\pm 0.09\%$ \\
\hline
\end{tabular}
\EC
\renewcommand{\arraystretch}{1}
\caption{Examples of EWPO with their current absolute and relative
experimental errors (see text).}
\label{tab:ewpotoday}
\end{table}

The prospective accuracy that can be achieved for electroweak precision
observables at the next generation of colliders, including $\mt$ and
$\mh$, has been analyzed in detail in \citere{blueband} and is reviewed in 
\refta{tab:ewpofut}. 
%
\begin{table}[htb]
\renewcommand{\arraystretch}{1.5}
\BC
\begin{tabular}{|c||c||c|c|c|c|c|}
\cline{2-7} \multicolumn{1}{c||}{}
& now & Tev.\ Run~IIA & Run~IIB & LHC & ~LC~  & GigaZ \\
\hline\hline
$\de\sweff(\times 10^5)$ & 17   & 78   & 29   & 14--20 & (6)  & 1.3  \\
\hline
$\de\MW$ [MeV]           & 34   & 27   & 16   & 15   & 10   & 7      \\
\hline
$\de\mt$ [GeV]           &  5.1 &  2.7 &  1.4 &  1.0 &  0.2 & 0.13   \\
\hline
$\de\mh$ [MeV]           &  --- &  --- & ${\cal O}(2000)$ 
                                                  &  100 &   50 &   50 \\
\hline
\end{tabular}
\EC
\renewcommand{\arraystretch}{1}
\caption{Current and anticipated future experimental uncertainties for 
$\sweff$, $\MW$, $\mt$, and $\mh$. 
See \citere{blueband} for a detailed discussion and further references.}
\label{tab:ewpofut}
\end{table}


\section{Theory status of precision observables in the MSSM}

In this section we discuss the theory status of the various EWPO in the
MSSM and for sake of comparison also in the SM. In order to analyze
virtual effects of SUSY, it is in general not sufficient to restrict to
certain parameterizations, like the $S$, $T$, $U$ parameters~\cite{stu}
(which are only applicable for specific types of new physics
contributions and are intrinsically one-loop quantities; for a
discussion of this issue, see \citere{orangebookstu}). Instead, the MSSM
predictions for the actual observables need to be worked out in detail.

Concerning the situation in the SM, as will be described in detail below, 
the level of accuracy for EWPO is quite advanced. Obtaining predictions
for observables in the MSSM at a certain order requires in general a
higher effort than for the SM case. This is related to the fact that in
the MSSM many additional parameters enter, in particular new mass
scales. The level of accuracy achieved so far in the MSSM is therefore
somewhat lower than in the SM.

Furthermore, besides the known sources of sizable corrections in the
SM, e.g.\ contributions enhanced by powers of $\mt$ or logarithms of
light fermions, there are additional sources of possibly large 
corrections within the MSSM:

\begin{itemize}

\item
Large corrections can arise not only from loops containing the top
quark, but also its scalar superpartners. Corrections from the top and
scalar top quark sector of the MSSM can be especially large in the MSSM
Higgs sector, where one-loop corrections can reach the level of 100\%.
The leading one-loop term from the top and scalar top sector entering
the predictions in the Higgs sector is given by~\cite{ERZ}
\BE
\sim \gf \, \mt^4 \log\KL \frac{\mste\mstz}{\mt^2} \KR~.
\EE

\item
Effects from the $b/\Sbot$ sector of the MSSM can also be very important
for large $\tb$.

\item
The $b$~Yukawa coupling can receive large SUSY corrections, yielding a
shift in the relation between the $b$~quark mass and the corresponding
Yukawa coupling~\cite{deltamb1},
\BE
y_b = \frac{\wz}{v\,\Cb} \frac{\mb}{1 + \De\mb}~.
\EE
The quantity $\De\mb$ contains in particular a contribution involving a
gluino in the loop, which gives rise to a correction proportional to
$(\als \,\mu\,\mgl\,\tb)$, which can be large. For $\De\mb \to -1$ the 
$b$~Yukawa coupling even becomes non-perturbative.

\item
In general, SUSY loop contributions can become large if some of the SUSY
particles are relatively light.

\end{itemize}


\subsection{Electroweak precision observables}

Within the SM, very accurate results are in particular available for
$\MW$, where meanwhile all ingredients of the complete two-loop result
are known~\cite{mwtwoloop,MWferm2} (as well as leading QCD and electroweak 
three-loop corrections). Taking into account the latest result obtained in
\citere{delrho3lew}, the remaining theoretical uncertainties from unknown
higher-order corrections within the SM can be estimated to
be (using the methods described in \citeres{MWferm2,blueband,radcor02})
\BE
{\rm SM:} \quad \de \MW^{\rm th} \approx \pm 4 \mev, \quad
\de \sweff^{\rm th} \approx \pm 6 \times 10^{-5}.
\label{eq:unchighord}
\EE
They are considerably smaller at present than the parametric uncertainties
from the experimental errors of the input parameters $\mt$ and
$\De\al_{\rm had}$. The experimental errors of 
$\de \mt = \pm 5.1$~GeV and
$\de(\De\al_{\rm had}) = 36 \times 10^{-5}$~\cite{blueband_s02}
induce parametric theoretical uncertainties of
\BEA
\de\mt: \quad
\de \MW^{\rm para} \approx \pm 31 \mev, && 
     \de \sweff^{\rm para} \approx \pm 16 \times 10^{-5}, \non \\
\de(\De\al_{\rm had}): \quad
\de \MW^{\rm para} \approx \pm 6.5 \mev, && 
     \de \sweff^{\rm para} \approx \pm 13 \times 10^{-5}. 
\EEA
This has to be compared with the current experimental errors given in
\refta{tab:ewpotoday}. 

At one-loop order, complete results for the electroweak precision
observables $\MW$ and $\sweff$ are also known within the
MSSM. At the two-loop level, the leading
corrections in $\oaas$ have been obtained~\cite{dr2lA}, which enter via the 
quantity $\De\rho$,
\BE
\De\rho = \frac{\Si_Z(0)}{\MZ^2} - \frac{\Si_W(0)}{\MW^2} .
\label{delrho}
\EE
It parameterises the leading universal corrections to the electroweak
precision observables induced by
the mass splitting between fields in an isospin doublet~\cite{rho}.
$\Si_{Z,W}(0)$ denote the transverse parts of the unrenormalized $Z$-
and $W$-boson self-energies at zero momentum transfer, respectively.
The induced shifts in $\MW$ and $\sweff$ are in leading order given by
(with $1-\sw^2 \equiv \cw^2 = \MW^2/\MZ^2$)
\BE
\de\MW \approx \frac{\MW}{2}\frac{\cw^2}{\cw^2 - \sw^2} \De\rho, \quad
\de\sweff \approx - \frac{\cw^2 \sw^2}{\cw^2 - \sw^2} \De\rho .
\label{precobs}
\EE
For the gluonic corrections, results in $\oaas$ have also been obtained
for the prediction of $\MW$~\cite{dr2lB}. The comparison with the
contributions entering via $\De\rho$ showed that in this case indeed the
full result is well approximated by the $\De\rho$ contribution. Contrary
to the SM case, the \twol\ $\oaas$ corrections turned out to increase
the \onel\ contributions, leading to an enhancement of up to
35\%~\cite{dr2lA}.

\begin{figure}[ht!]
\includegraphics[width=7.5cm,height=6.0cm]{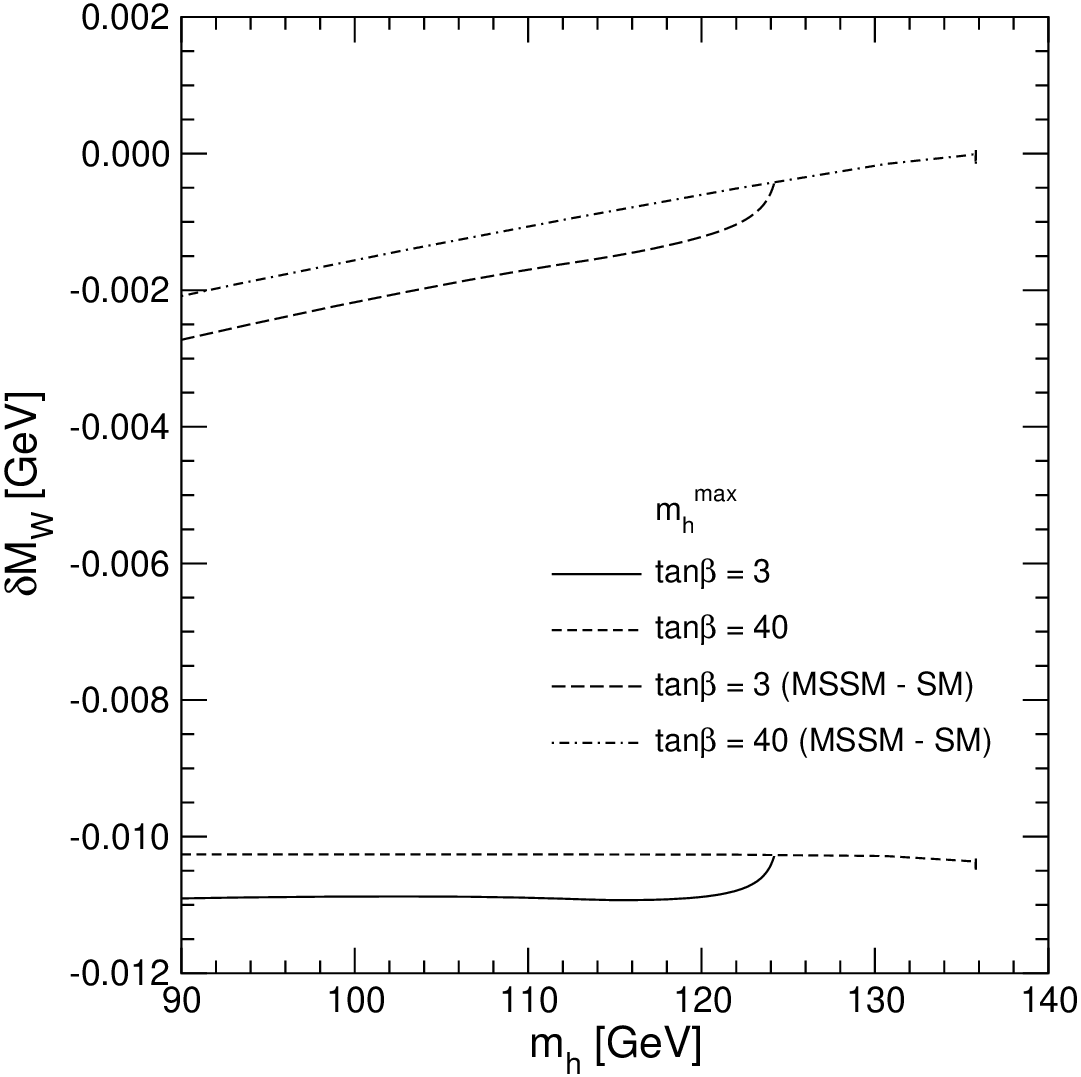}
\includegraphics[width=7.5cm,height=6.0cm]{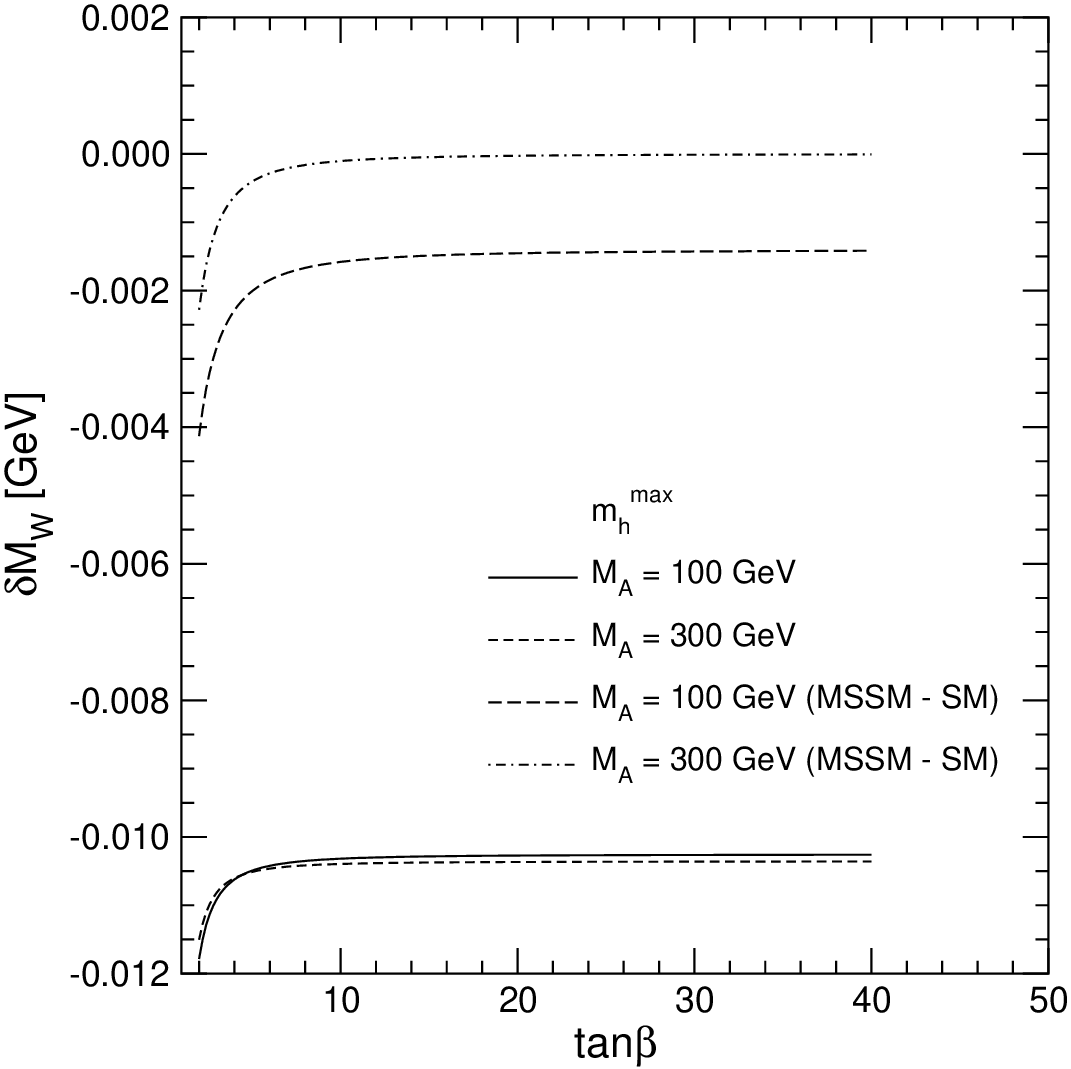}
\caption{Contribution of the \order{\al_t^2} MSSM corrections to $\MW$
as a function of $\mh$ (left) and $\tan\be$ (right) in the
$\mhmax$~scenario~\cite{benchmark2}.}
\label{fig:delMWMSSM}
\end{figure}

Recently the leading \twol\ corrections to 
$\De\rho$ at \order{\al_t^2}, \order{\al_t \al_b}, \order{\al_b^2}
($\al_{t, b} \equiv y_{t, b}^2 / (4 \pi)$, $y_{t, b}$ being the 
top and bottom Yukawa couplings, respectively)
have been obtained for the case of a large SUSY scale, 
$\msusy \gg \MZ$~\cite{dr2lal2mh0,dr2lal2}. These contributions involve
the top and bottom Yukawa couplings and
contain in particular corrections proportional to $\mt^4$ and bottom
loop corrections enhanced by $\tb$. As an example, the effect of the
\order{\al_t^2} MSSM contributions on $\de\MW$ amounts up to $-12 \mev$,
see \reffi{fig:delMWMSSM}. The `effective' change in $\MW$ in comparison
with the corresponding SM result with
the same value of the Higgs-boson mass is significantly smaller. It
amounts up to $-3 \mev$ and goes to zero for large $\MA$ as expected from the
decoupling behavior.

Comparing the presently available results for the electroweak precision
observables $\MW$ and $\sweff$ in the MSSM with those in the SM (as
given in \refeq{eq:unchighord}), a crude estimate yields
\BE
{\rm MSSM:} \quad \de \MW^{\rm th} \approx \pm 10 \mev, \quad
\de \sweff^{\rm th} \approx \pm 12 \times 10^{-5}.
\label{eq:unchighordmssm}
\EE
Thus, the uncertainties from unknown higher-order corrections within
the MSSM are about twice as large as in the SM in the case of $\sweff$
and even larger for $\MW$.


\subsection{The lightest $\cp$-even Higgs boson mass}

The mass of the lightest $\cp$-even MSSM Higgs boson can be predicted from 
the other model parameters. At the tree-level, the two $\cp$-even Higgs 
boson masses are obtained by rotating the neutral $\cp$-even Higgs boson
mass matrix with an angle $\alpha$, 
\BEA M_{\rm Higgs}^{2, {\rm tree}} &=& \ML \MA^2 \SQb +
\MZ^2 \CQb & -(\MA^2 + \MZ^2) \Sb \Cb \\ -(\MA^2 + \MZ^2) \Sb \Cb &
\MA^2 \CQb + \MZ^2 \SQb \MR ,
\label{higgsmassmatrixtree}
\EEA
with  $\alpha$ satisfying
\BE
\tan 2\alpha = \tan 2\beta \frac{\MA^2 + \MZ^2}{\MA^2 - \MZ^2},
\quad - \frac{\pi}{2} < \alpha < 0 . 
\label{alpha}
\EE
In the Feynman-diagrammatic approach the higher-order corrected 
Higgs boson masses are derived by finding the
poles of the $h,H$-propagator 
matrix whose inverse is given by
\BE
\left(\Delta_{\rm Higgs}\right)^{-1}
= - i \ML p^2 -  m_{H, \rm tree}^2 + \hSi_{HH}(p^2) &  \hSi_{hH}(p^2) \\
     \hSi_{hH}(p^2) & p^2 -  m_{h, \rm tree}^2 + \hSi_{hh}(p^2) \MR,
\label{higgsmassmatrixnondiag}
\EE
where the $\hSi(p^2)$ denote the renormalized Higgs-boson self-energies,
$p$ being the momentum going through the external legs.
Determining the poles of the matrix $\Delta_{\rm Higgs}$ in
\refeq{higgsmassmatrixnondiag} is equivalent to solving
the equation
\begin{equation}
\left[p^2 - m_{h, \rm tree}^2 + \hSi_{hh}(p^2) \right]
\left[p^2 - m_{H, \rm tree}^2 + \hSi_{HH}(p^2) \right] -
\left[\hSi_{hH}(p^2)\right]^2 = 0\,.
\label{eq:proppole}
\end{equation}

The status of the available results for the self-energy contributions to
\refeq{higgsmassmatrixnondiag} can be summarized as follows. For the
one-loop part, the complete result within the MSSM is 
known~\cite{ERZ,mhiggsf1lB,mhiggsf1lC}. The by far dominant
one-loop contribution is the \order{\al_t} term due to top and stop 
loops. Concerning the two-loop
effects, their computation is quite advanced and it has now reached a
stage such that all the presumably dominant
contributions are known, see \citere{mhiggsAEC,spmartin} and references
therein. They include the strong corrections, usually 
indicated as \order{\al_t\als}, and Yukawa corrections, \order{\al_t^2},
to the dominant one-loop \order{\al_t} term, as well as the strong
corrections to the bottom/sbottom one-loop \order{\al_b} term,
i.e.\ the \order{\al_b\als} contribution. 
For the $b/\Sbot$~sector
corrections also an all-order resummation of the $\Tb$-enhanced terms,
\order{\al_b(\als\tb)^n}, is known.
Most recently the \order{\al_t \al_b} and \order{\al_b^2} corrections
have been derived~\cite{mhiggsEP5}.
All \twol\ corrections have been obtained by neglecting the external
momentum.

An upper bound of $\mh \lsim 135 \gev$~\cite{mhiggslong,mhiggsAEC} can
be established~\cite{feynhiggs} taking into 
account all existing higher-order corrections (in the
$\mhmax$~scenario with $\mt = 174.3$~GeV and $\msusy = 1$~TeV, see 
\citeres{benchmark,benchmark2}).

The remaining theoretical higher-order uncertainties in $\mh$ have
been analyzed in detail in \citere{mhiggsAEC}. This has been done by
\begin{itemize}
\item
extrapolating from the size of the existing \onel\ corrections to the missing
\twol\ contributions,
\item
changing the renormalization scale in the \onel\
result~\cite{feynhiggs1.2} in order to estimate missing \twol\ corrections
\item
changing the renormalization of the top quark mass at the \twol\ level
in order to estimate remaining three-loop contributions,
\item
from the result for the leading three-loop contribution.
\end{itemize}
As a result, the remaining theoretical uncertainty from unknown higher
orders has been estimated to be
\BE
\de\mh^{\rm th} \approx \pm 3 \gev~.
\EE

Concerning the parametric uncertainties from the experimental errors of
the input parameters, in particular the current experimental error of
the top-quark mass of $\de\mt = \pm 5.1$~GeV has a very large 
effect~\cite{Heinemeyer:1999zf},
\BE
\de\mt: \quad
\de \mh^{\rm para} \approx \pm 5 \gev~.
\EE

In order to enable sensitive electroweak precision tests in the MSSM
Higgs sector a drastic reduction of the parametric uncertainty induced
by $\de\mt$ will be crucial~\cite{deltamt}. This can be achieved with the 
measurement of $\mt$ at the LC~\cite{teslatdr,nlc,jlc},
\BE
\de\mt^{\rm exp, LC} \lsim 100 \mev~.
\EE
Besides a drastically improved experimental precision on $\mt$,
obviously also a big reduction of the theoretical uncertainties from
unknown higher-order corrections will be necessary. In 
order to match the future precision on $\mh$ with the accuracy of the
theoretical prediction, the theoretical error has to be
reduced by at least a factor of~10. This will require a complete \twol\
calculation, including the external momentum, dominant three-loop and
possibly even leading four-loop corrections.


\subsection{$B \to X_s \ga$ and the anomalous magnetic moment of the
muon}

Examples of further observables where virtual effects of SUSY particles
can be very important are the branching ratio for $B \to X_s \ga$ and 
the anomalous magnetic moment of the muon, $g_\mu - 2$. 

The branching ratio $\br(b \to s \ga)$ can receive large SUSY
corrections for light charged Higgs bosons and large $\mu$ or $\tb$. The
flavour-changing neutral current processes impose very important
constraints on the parameter space both of general two-Higgs-doublet
models and of the MSSM. The currently available SUSY contributions to 
$\br(b \to s \ga)$ include the one-loop result and leading higher-order
corrections~\cite{bsgth}.

SUSY contributions to $g_\mu - 2$ are particularly important for large
$\tb$ and light gaugino and slepton masses. The one-loop result in the
MSSM has been supplemented by leading logarithmic two-loop
contributions~\cite{Degrassi:1998es}. The comparison of the theoretical
predictions with the experimental result~\cite{gm2} is affected
by sizable QCD uncertainties, see \citere{gm2disc} for a discussion.


\section{Precision tests of the SM and the MSSM}



Before investigating the case of the unconstrained MSSM, we first
focus on the example of mSUGRA as a particular SUSY-breaking
scenario. It is interesting to note that the rather restricted scenario of
mSUGRA is still compatible with all available constraints from EWPO, the
Higgs boson sector, cold dark matter (CDM)~\cite{cdm}, $\br(b \to s
\ga)$~\cite{bsg} and from the anomalous magnetic moment of the muon, 
$g_\mu - 2$~\cite{gm2}, see e.g.\ \citeres{ellisnew,ehow2}.
The constraints from $\br(b \to s \ga)$ and in particular $g_\mu - 2$
favor the positive sign of the parameter $\mu$, see e.g.\
\citere{gm2mSUGRA}. The prediction for $g_\mu - 2$, however, is still 
affected by sizable QCD uncertainties, as discussed above.

\begin{figure}[htb!]
\begin{center}
\epsfig{figure=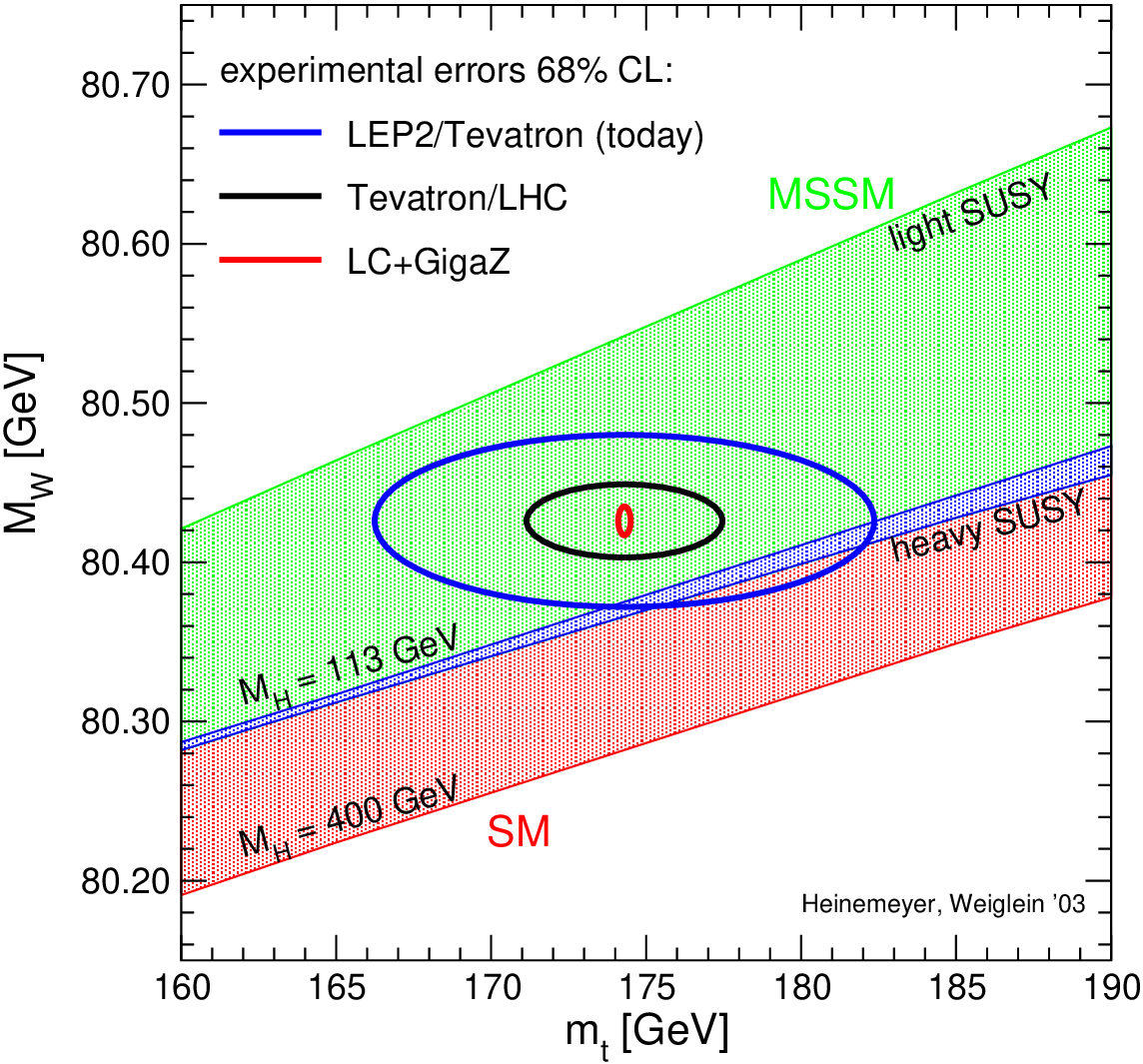,width=12cm}
\end{center}
\caption[]{The current experimental results for $\MW$ and $\mt$ and the
prospective accuracies at the next generation of colliders are shown in
  comparison with the SM prediction (red and blue bands) and the MSSM
prediction (green and blue bands).
}
\label{fig:MWMT}
\end{figure}

We now turn to the unconstrained MSSM and compare it with the EWPO data. 
In \reffi{fig:MWMT} we compare the SM and the MSSM prediction for $\MW$
as a function of $\mt$~\cite{MWMTplot}. The predictions within the two models 
give rise to two bands in the $\mt$--$\MW$ plane with only a relatively small
overlap region (indicated by a blue area in \reffi{fig:MWMT}). 
The allowed parameter region in the SM (the red
and blue bands) arises from varying the only free parameter of the
model, the mass of the SM Higgs boson, from $\MH = 113 \gev$ (upper edge
of the blue area) to $400 \gev$ (lower edge of the red area).
The green and the blue areas indicate
the allowed region for the unconstrained MSSM. SUSY masses close
to their experimental lower limit are assumed for the upper edge of the green
area, while the decoupling limit with SUSY masses of \order{2 \tev}
yields the lower edge of the blue area. Thus, the overlap region between
the predictions of the two models corresponds in the SM to the region
where the Higgs boson is light, i.e.\ in the MSSM allowed region ($\mh
\lsim 135 \gev$). In the MSSM it corresponds to the case where all
superpartners are heavy, i.e.\ the decoupling region of the MSSM. 
The current 68\%~C.L.\ experimental results for $\mt$ and $\MW$ slightly favor
the MSSM over the SM. 
The prospective accuracies for the LHC and the LC with GigaZ option,
see \refta{tab:ewpofut}, are also shown in the plot (using the current
central values), indicating the
potential for a significant improvement of the sensitivity of the
electroweak precision tests~\cite{gigaz}.

\begin{figure}[htb!]
\begin{center}
\epsfig{figure=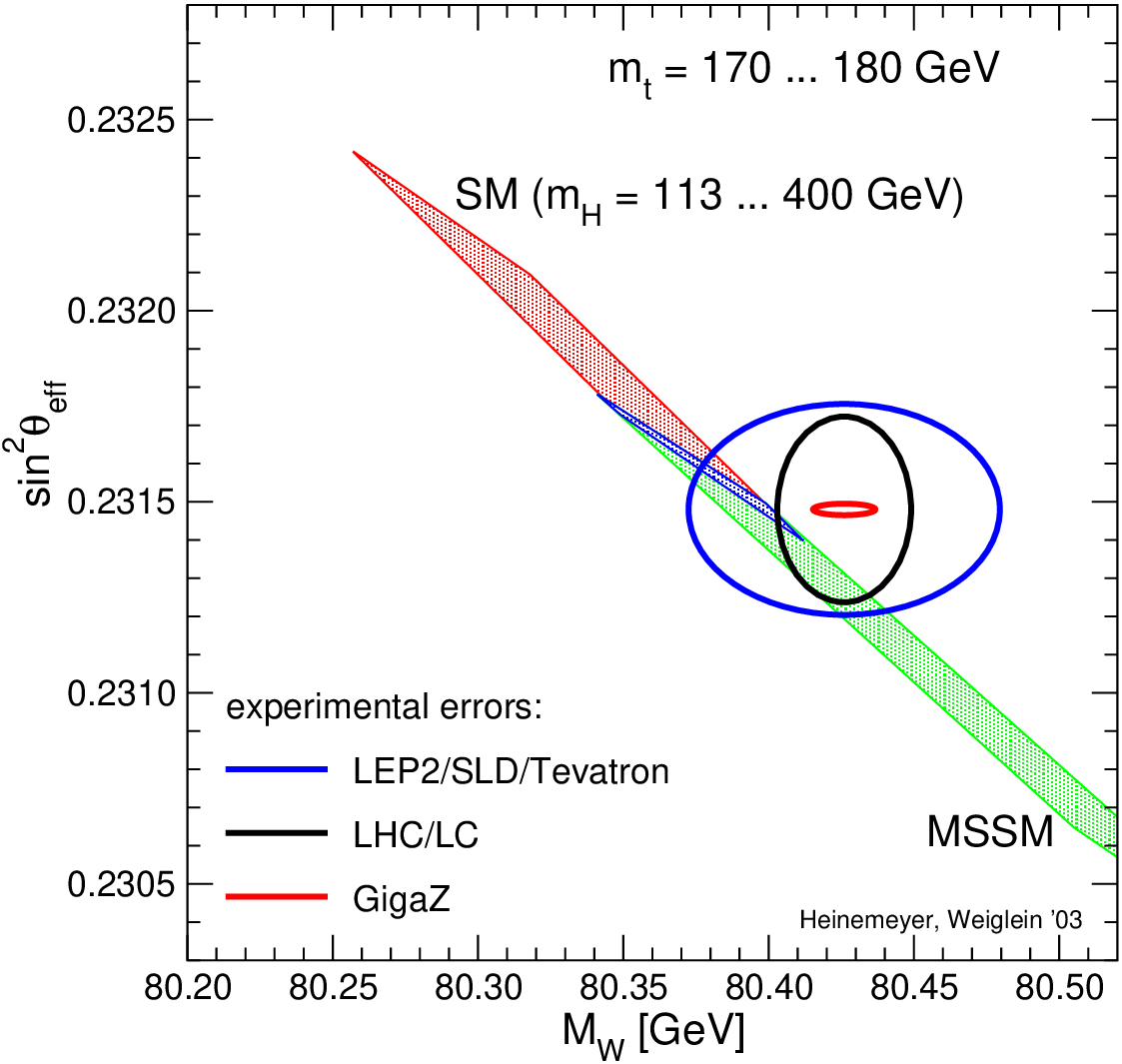,width=12cm}
\end{center}
\caption[]{The current experimental results for $\MW$ and $\sweff$ and the
prospective accuracies at the next generation of colliders are shown in
  comparison with the SM prediction (red and blue bands) and the MSSM
prediction (green and blue bands).
}
\label{fig:MWsw2eff}
\end{figure}

In \reffi{fig:MWsw2eff} the comparison between the SM and the MSSM is
shown in the $\MW$--$\sweff$ plane. As above, the predictions in the SM
(red and blue bands) and the MSSM (green and blue bands) are shown
together with the current 68\%~C.L.\ experimental results and the 
prospective accuracies for the LHC and the LC with GigaZ option. Again
the MSSM is slightly favored over the SM. It should be noted that the
prospective improvements in the experimental accuracies, in particular
at a LC with GigaZ option, will provide a high sensitivity to deviations
both from the SM and the MSSM.

The central value for the experimental value of $\sweff$ in
\reffi{fig:MWsw2eff} is based on
both leptonic and hadronic data. The fact that the two most precise
measurements, $A_{\rm LR}$ from SLD~\cite{alr} and $A^{\rm b}_{\rm FB}$ from 
LEP~\cite{afb}, differ from each other by about $3\si$, giving rise to a
relatively low fit probability of the SM global fit, has caused
considerable attention in the literature. In particular, several
analyses have been performed where the hadronic data on $A_{\rm FB}$
have been excluded from the global fit 
(see e.g.\ \citeres{chanowitz,lightsf}). It has been noted that in this
case the SM global fit, possessing a much higher fit probability, yields
an upper bound on $\MH$ which is rather low in view of the experimental
lower bound on $\MH$ of $\MH > 114.4$~GeV~\cite{mhLEPfinal}. The value of 
$\sweff$ corresponding to the measurement of $A_{\rm LR}({\rm SLD})$
alone is $\sweff = 0.23098 \pm 0.00026$~\cite{alr}. \reffi{fig:MWsw2eff}
shows that adopting the latter value of $\sweff$ makes the agreement between
the data and the SM prediction much worse, while the MSSM provides a
very good description of the data. In accordance with this result, in
\citere{lightsf} it has been found that the contribution of light
gauginos and scalar leptons in the MSSM (in a scenario with vanishing
SUSY contribution to $\De\rho$) gives rise to a shift in $\MW$
and $\sweff$ as compared to the SM case which brings the MSSM prediction
in agreement with the experimental values of $\MW$ and $A_{\rm LR}({\rm SLD})$.

On the other hand, it has also been investigated whether the discrepancy
between $A_{\rm LR}$ and $A^{\rm b}_{\rm FB}$ could be explained in
terms of contributions of some kind of new physics. The (loop-induced)
contributions from SUSY particles in the MSSM are however too small to
account for the $3\si$ difference between the two observables (see e.g.\
\citere{lightsf}). Thus, the quality of the fit to $A_{\rm LR}$ and
$A^{\rm b}_{\rm FB}$ in the MSSM is similar to the one in the SM.


Another observable for which the SM prediction shows a large deviation
by about $3 \si$ from the experimental value is the neutrino--nucleon 
cross section measured at NuTeV~\cite{nutev}. Also in this case loop
effects of SUSY particles in the MSSM are too small to account for a
sizable fraction of the discrepancy (see e.g.\ \citere{Davidson:2001ji}).

\begin{figure}[htb!]
\begin{center}
\epsfig{figure=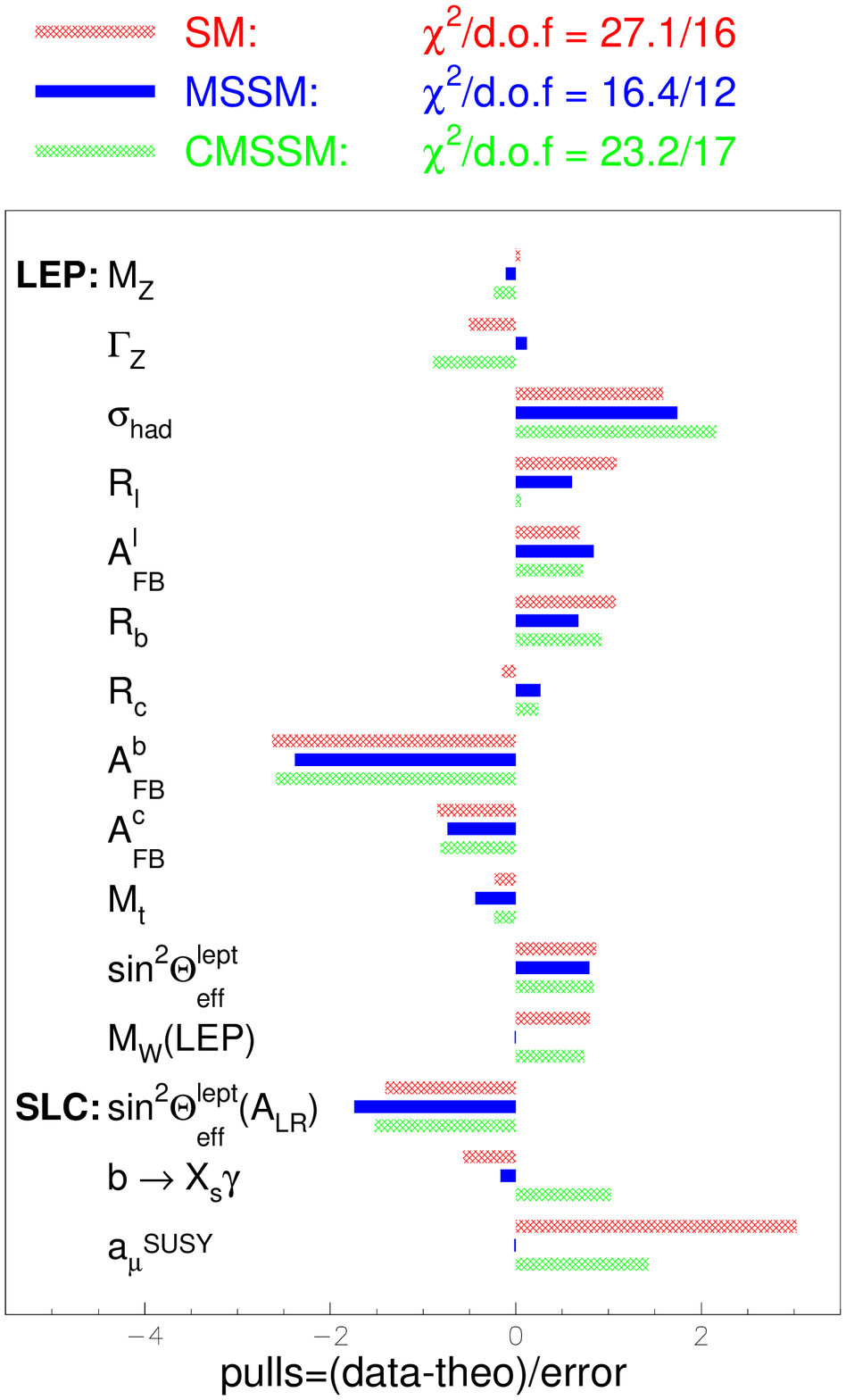,width=12cm,height=15cm}
\end{center}
\caption[]{The predictions in the SM, the MSSM and the mSUGRA scenario
(CMSSM) are compared with the data. Deviations between theory and
experiment are indicated in units of one standard deviation of the
experimental results (from \citere{ewpoMSSM}).
}
\label{fig:ewpoMSSM}
\end{figure}
 
A global fit to all data has been performed within the MSSM in
\citere{ewpoMSSM}. The results are shown in \reffi{fig:ewpoMSSM}, where
the predictions in the SM, the MSSM and the constrained MSSM (i.e.\ the
mSUGRA scenario) are compared with the experimental data (the SUSY
predictions are for $\tan\be = 35$).
\reffi{fig:ewpoMSSM} shows the features discussed above: the MSSM
predictions for $\MW$ and (for large $\tan\be$) $g_\mu - 2$ are in better 
agreement with the
data than in the SM (slight improvements also occur for the
total width of the $Z$~boson, $\Gamma_Z$, and for $B \to X_s \ga$).
On the other hand, for the observables with the largest deviations
between theory and experiment, namely $A^{\rm b}_{\rm FB}$ and the
neutrino--nucleon cross section measured at NuTeV (the latter is not
shown in \reffi{fig:ewpoMSSM}), the MSSM does not yield a significant
improvement compared to the SM. The global fit in the MSSM has a lower
$\chi^2$ value than in the SM. Since the MSSM fit has less degrees of
freedom than the SM one, the overall fit probability in the MSSM is only
slightly better than in the SM.


\section{Conclusions}

We have investigated electroweak precision tests in the framework of the
MSSM. Compared to the SM, the MSSM contains several new sources for
potentially large radiative corrections. Of particular importance is the
Higgs sector of the MSSM. While within the SM the Higgs-boson mass is a
free parameter, the relation between the mass of the lightest $\cp$-even
Higgs boson of the MSSM and the other model parameters is one of the
most striking predictions of SUSY models. 

We have summarized the theory status of the precision observables in the 
MSSM and have given an estimate of the remaining theoretical uncertainties 
from unknown higher-order corrections. We find that the present theoretical 
uncertainties for $\MW$ and the $Z$-boson observables in the MSSM are 
still significantly higher than in the SM.

We have discussed examples of electroweak precision tests in the
context of the mSUGRA scenario and the unconstrained MSSM. The mSUGRA 
scenario, despite its small number of free parameters, can accommodate
all experimental constraints, i.e.\ the ones from the electroweak
precision data, the anomalous magnetic moment of the muon, 
$B \to X_s \ga$, the cold dark matter constraints, 
and the lower bounds from the Higgs and SUSY particle searches.
The global fit to all data in the MSSM yields a better agreement for
$\MW$ and $g_\mu - 2$ than in the SM, while no significant improvements
occur for $A^{\rm b}_{\rm FB}$ and the neutrino--nucleon cross section 
measured at NuTeV. The overall fit probability in the MSSM is only
slightly better than the one in the SM.

We have furthermore analyzed the potential of anticipated future
electroweak precision data, including the Higgs boson and the top-quark
mass, for testing quantum effects of the MSSM. In order to match the
prospective accuracy of $\mh$ achievable at the LHC with the theoretical
prediction, an improvement of the theoretical uncertainties from unknown
higher-order corrections by more than a factor ten will be necessary. 
In order to reduce the parametric theoretical uncertainties to the same
level, the precision measurement of the top-quark mass at the LC will be
crucial. Significant improvements of both the parametric uncertainties
and the ones from unknown higher-order corrections will also be
necessary in view of the prospective experimental accuracies of $\MW$
and $\sweff$ at the next generation of colliders. Thus, substantial
progress in the theoretical predictions will be necessary in order
to exploit electroweak precision physics in the MSSM, which might become
possible at the next generation of colliders.



\end{document}

%% file: paperdef.tex
\makeatletter
\def\citer{\@ifnextchar [{\@tempswatrue\@citexr}{\@tempswafalse\@citexr[]}}
 
\def\@citexr[#1]#2{\if@filesw\immediate\write\@auxout{\string\citation{#2}}\fi
  \def\@citea{}\@cite{\@for\@citeb:=#2\do
    {\@citea\def\@citea{--\penalty\@m}\@ifundefined
       {b@\@citeb}{{\bf ?}\@warning
       {Citation `\@citeb' on page \thepage \space undefined}}%
\hbox{\csname b@\@citeb\endcsname}}}{#1}}
\makeatother

\def\refeq#1{\mbox{Eq.~(\ref{#1})}}

\def\reffi#1{\mbox{Fig.~\ref{#1}}}

\def\refta#1{\mbox{Tab.~\ref{#1}}}
\def\citere#1{\mbox{Ref.~\cite{#1}}}
\def\citeres#1{\mbox{Refs.~\cite{#1}}}

\def\xtilde#1{%
  \setbox0\hbox{$\tilde#1$}%
  \rlap{\raise\ht0\hbox{\tiny$_{\,(\;\,)}$}}%
  \tilde#1%
}

\newcommand{\mste}{m_{\tilde{t}_1}}
\newcommand{\mstz}{m_{\tilde{t}_2}}

\newcommand{\msusy}{M_{\text{SUSY}}}

\newcommand{\oaas}{{\cal O}(\alpha\alpha_s)}

\def\order#1{${\cal O}(#1)$}
\newcommand{\cp}{{\cal CP}}

\newcommand{\wz}{\sqrt{2}}

\newcommand{\twol}{two-loop}
\newcommand{\onel}{one-loop}

\newcommand{\MW}{M_W}
\newcommand{\MZ}{M_Z}
\newcommand{\MA}{M_A}
\newcommand{\mh}{m_h}

\newcommand{\MH}{M_H}

\newcommand{\mhmax}{m_h^{\rm max}}

\newcommand{\mt}{m_{t}}

\newcommand{\mb}{m_{b}}

\newcommand{\mgl}{m_{\tilde{g}}}

\newcommand{\Sbot}{\tilde{b}}

\newcommand{\tsf}{\theta\kern-.20em_{\tilde{f}}}
\newcommand{\tsfp}{\theta\kern-.20em_{\tilde{f}\prime}}
\newcommand{\tsq}{\theta\kern-.15em_{\tilde{q}}}
\newcommand{\sw}{s_W}
\newcommand{\cw}{c_W}
\newcommand{\sweff}{\sin^2\theta_{\mathrm{eff}}}

\newcommand{\KL}{\left(}
\newcommand{\KR}{\right)}

\newcommand{\VL}{\left( \begin{array}{c}}
\newcommand{\VR}{\end{array} \right)}
\newcommand{\ML}{\left( \begin{array}{cc}}
\newcommand{\MLd}{\left( \begin{array}{ccc}}
\newcommand{\MLv}{\left( \begin{array}{cccc}}
\newcommand{\MR}{\end{array} \right)}

\newcommand{\Tb}{\tan \beta\hspace{1mm}}
\newcommand{\tb}{\tan \beta}

\newcommand{\Sb}{\sin \beta}
\newcommand{\SQb}{\sin^2\beta\hspace{1mm}}

\newcommand{\Cb}{\cos \beta\hspace{1mm}}
\newcommand{\CQb}{\cos^2\beta\hspace{1mm}}

\newcommand{\tev}{\,\, \mathrm{TeV}}
\newcommand{\gev}{\,\, \mathrm{GeV}}
\newcommand{\mev}{\,\, \mathrm{MeV}}

\newcommand{\BC}{\begin{center}}
\newcommand{\EC}{\end{center}}
\newcommand{\BE}{\begin{equation}}
\newcommand{\EE}{\end{equation}}
\newcommand{\BEA}{\begin{eqnarray}}
\newcommand{\BEAnn}{\begin{eqnarray*}}
\newcommand{\EEA}{\end{eqnarray}}
\newcommand{\EEAnn}{\end{eqnarray*}}
\newcommand{\non}{\nonumber}
\newcommand{\id}{{\rm 1\kern-.12em
\rule{0.3pt}{1.5ex}\raisebox{0.0ex}{\rule{0.1em}{0.3pt}}}}
\newcommand{\lsim}
{\;\raisebox{-.3em}{$\stackrel{\displaystyle <}{\sim}$}\;}

\newcommand{\gf}{G_F}

\def\al{\alpha}

\def\als{\alpha_s}

\def\be{\beta}

\def\ga{\gamma}

\def\de{\delta}

\def\si{\sigma}

\def\Ga{\Gamma}

\def\De{\Delta}

\def\Si{\Sigma}

\def\hSi{\hat{\Sigma}}

\newcommand{\epem}{$e^+e^-$}

\newcommand{\br}{{\rm BR}}



%% file: susyewpo_hepph.bbl
\begin{thebibliography}{99}
\frenchspacing

\bibitem{susy} 
H.~P.~Nilles,
Phys.\ Rept.\  {\bf 110} (1984) 1;\\
H.~E.~Haber and G.~L.~Kane,
Phys.\ Rept.\  {\bf 117} (1985) 75;\\
R.~Barbieri,
Riv.\ Nuovo Cim.\  {\bf 11N4} (1988) 1.

\bibitem{pdg} 
K.~Hagiwara {\it et al.}  [Particle Data Group Collaboration],
Phys.\ Rev.\ D {\bf 66} (2002) 010001.

\bibitem{hhg} 
J.~Gunion, H.~Haber, G.~Kane and S.~Dawson,
{\em The Higgs Hunter's Guide}, Addison-Wesley, 1990.

\bibitem{mhiggslong}
S.~Heinemeyer, W.~Hollik and G.~Weiglein,
Eur.\ Phys.\ J.\ C {\bf 9} (1999) 343
[arXiv:hep-ph/9812472].

\bibitem{mhiggsAEC}
G.~Degrassi, S.~Heinemeyer, W.~Hollik, P.~Slavich and G.~Weiglein,
Eur.\ Phys.\ J.\ C {\bf 28} (2003) 133
[arXiv:hep-ph/0212020].

\bibitem{lhc}
 ATLAS Collaboration, Technical Design Report 1999, Vol.~II,
 CERN/LHC/99-15, ATLAS TDR 15.

\bibitem{teslatdr}
J.~A.~Aguilar-Saavedra {\it et al.}  [ECFA/DESY LC Physics Working Group
                  Collaboration],
arXiv:hep-ph/0106315.

\bibitem{nlc}
T.~Abe {\it et al.}  [American Linear Collider Working Group Collaboration],
in {\it Proc. of the APS/DPF/DPB Summer Study on the Future of
  Particle Physics (Snowmass 2001) } ed. N.~Graf, 
arXiv:hep-ex/0106056.

\bibitem{jlc}
K.~Abe {\it et al.}  [ACFA Linear Collider Working Group Collaboration],
arXiv:hep-ph/0109166.

\bibitem{benchmark}
M.~Carena, S.~Heinemeyer, C.~E.~Wagner and G.~Weiglein,
arXiv:hep-ph/9912223.

\bibitem{benchmark2}
M.~Carena, S.~Heinemeyer, C.~E.~Wagner and G.~Weiglein,
Eur.\ Phys.\ J.\ C {\bf 26} (2003) 601
[arXiv:hep-ph/0202167].

\bibitem{sps}
B.~C.~Allanach {\it et al.},
in {\it Proc. of the APS/DPF/DPB Summer Study on the Future of
  Particle Physics (Snowmass 2001) } ed. N.~Graf, 
Eur.\ Phys.\ J.\ C {\bf 25} (2002) 113
[eConf {\bf C010630} (2001) P125]
[arXiv:hep-ph/0202233].


\bibitem{gmsbrev}
G.~F.~Giudice and R.~Rattazzi,
Phys.\ Rept.\  {\bf 322} (1999) 419
[arXiv:hep-ph/9801271].

\bibitem{amsb1}
L.~Randall and R.~Sundrum,
Nucl.\ Phys.\ B {\bf 557} (1999) 79
[arXiv:hep-th/9810155].

\bibitem{amsb2}
G.~F.~Giudice, M.~A.~Luty, H.~Murayama and R.~Rattazzi,
JHEP {\bf 9812} (1998) 027
[arXiv:hep-ph/9810442].

\bibitem{amsb3}
T.~Gherghetta, G.~F.~Giudice and J.~D.~Wells,
Nucl.\ Phys.\ B {\bf 559} (1999) 27
[arXiv:hep-ph/9904378].

\bibitem{asbs}
S.~Ambrosanio, A.~Dedes, S.~Heinemeyer, S.~Su and G.~Weiglein,
Nucl.\ Phys.\ B {\bf 624} (2002) 3
[arXiv:hep-ph/0106255];\\
J.~R.~Ellis, S.~Heinemeyer, K.~A.~Olive and G.~Weiglein,
Phys.\ Lett.\ B {\bf 515} (2001) 348
[arXiv:hep-ph/0105061].

\bibitem{asbs2}
A.~Dedes, S.~Heinemeyer, S.~Su and G.~Weiglein,
arXiv:hep-ph/0302174.


\bibitem{blueband}
U.~Baur, R.~Clare, J.~Erler, S.~Heinemeyer, D.~Wackeroth, G.~Weiglein
and D.~R.~Wood, 
in {\it Proc. of the APS/DPF/DPB Summer Study on the Future of
  Particle Physics (Snowmass 2001) } ed. N.~Graf, 
eConf {\bf C010630} (2001) P122
[arXiv:hep-ph/0111314].

\bibitem{stu}
M.~E.~Peskin and T.~Takeuchi,
Phys.\ Rev.\ D {\bf 46} (1992) 381.

\bibitem{orangebookstu}
T.~Abe {\it et al.}  [American Linear Collider Working Group
Collaboration],
in {\it Proc. of the APS/DPF/DPB Summer Study on the Future of Particle
Physics (Snowmass 2001) } ed. N.~Graf,
arXiv:hep-ex/0106057.

\bibitem{ERZ} 
J.~R.~Ellis, G.~Ridolfi and F.~Zwirner,
Phys.\ Lett.\ B {\bf 257} (1991) 83;\\
Y.~Okada, M.~Yamaguchi and T.~Yanagida,
Prog.\ Theor.\ Phys.\  {\bf 85} (1991) 1;\\
H.~E.~Haber and R.~Hempfling,
Phys.\ Rev.\ Lett.\  {\bf 66} (1991) 1815.

\bibitem{deltamb1}
L.~J.~Hall, R.~Rattazzi and U.~Sarid,
Phys.\ Rev.\ D {\bf 50} (1994) 7048
[arXiv:hep-ph/9306309].

\bibitem{mwtwoloop}
A.~Freitas, W.~Hollik, W.~Walter and G.~Weiglein,
Phys.\ Lett.\ B {\bf 495} (2000) 338
[arXiv:hep-ph/0007091];\\
M.~Awramik and M.~Czakon,
Phys.\ Rev.\ Lett.\  {\bf 89} (2002) 241801
[arXiv:hep-ph/0208113];\\
A.~Onishchenko and O.~Veretin,
Phys.\ Lett.\ B {\bf 551} (2003) 111
[arXiv:hep-ph/0209010];\\
M.~Awramik and M.~Czakon,
arXiv:hep-ph/0305248.


\bibitem{MWferm2}
A.~Freitas, W.~Hollik, W.~Walter and G.~Weiglein,
Nucl.\ Phys.\ B {\bf 632} (2002) 189
[arXiv:hep-ph/0202131].

\bibitem{delrho3lew}
M.~Faisst, J.~H.~Kuhn, T.~Seidensticker and O.~Veretin,
arXiv:hep-ph/0302275.

\bibitem{radcor02}
A.~Freitas, S.~Heinemeyer and G.~Weiglein,
hep-ph/0212068.

\bibitem{blueband_s02}
M.~W.~Grunewald,
Nucl.\ Phys.\ Proc.\ Suppl.\  {\bf 117} (2003) 280
[arXiv:hep-ex/0210003].

\bibitem{dr2lA}
A.~Djouadi, P.~Gambino, S.~Heinemeyer, W.~Hollik, C.~Junger and G.~Weiglein,
Phys.\ Rev.\ Lett.\  {\bf 78} (1997) 3626
[arXiv:hep-ph/9612363];
Phys.\ Rev.\ D {\bf 57} (1998) 4179
[arXiv:hep-ph/9710438].

\bibitem{rho}
M.~J.~Veltman,
Nucl.\ Phys.\ B {\bf 123} (1977) 89.

\bibitem{dr2lB} 
S.~Heinemeyer, PhD thesis, 
Universit\"at Karlsruhe, Shaker Verlag, Aachen 1998,
see {\tt www-itp.physik.uni-karlsruhe.de/prep/phd/};\\
G.~Weiglein,
arXiv:hep-ph/9901317;\\
S.~Heinemeyer, W.~Hollik and G.~Weiglein,
{\em in preparation}.
                
\bibitem{dr2lal2mh0} 
S.~Heinemeyer and G.~Weiglein,
in {\it Proc. of the 5th International Symposium on Radiative
  Corrections (RADCOR 2000) } ed. Howard E. Haber, 
arXiv:hep-ph/0102317.

\bibitem{dr2lal2} 
S.~Heinemeyer and G.~Weiglein,
JHEP {\bf 0210} (2002) 072
[arXiv:hep-ph/0209305].

\bibitem{mhiggsf1lB} 
P.~H.~Chankowski, S.~Pokorski and J.~Rosiek,
Phys.\ Lett.\ B {\bf 286} (1992) 307;
Nucl.\ Phys.\ B {\bf 423} (1994) 437
[arXiv:hep-ph/9303309].

\bibitem{mhiggsf1lC} 
A.~Dabelstein,
Nucl.\ Phys.\ B {\bf 456} (1995) 25
[arXiv:hep-ph/9503443];
Z.\ Phys.\ C {\bf 67} (1995) 495
[arXiv:hep-ph/9409375].

\bibitem{spmartin} 
S.~P.~Martin,
Phys.\ Rev.\ D {\bf 66} (2002) 096001
[arXiv:hep-ph/0206136].

\bibitem{mhiggsEP5} 
A.~Dedes, G.~Degrassi and P.~Slavich,
arXiv:hep-ph/0305127.

\bibitem{feynhiggs}
S.~Heinemeyer, W.~Hollik and G.~Weiglein,
Comput.\ Phys.\ Commun.\  {\bf 124} (2000) 76
[arXiv:hep-ph/9812320];\\
The {\em FeynHiggs} code can be obtained from {\tt www.feynhiggs.de} .

\bibitem{feynhiggs1.2} 
M.~Frank, S.~Heinemeyer, W.~Hollik and G.~Weiglein,
arXiv:hep-ph/0202166.

\bibitem{Heinemeyer:1999zf}
S.~Heinemeyer, W.~Hollik and G.~Weiglein,
JHEP {\bf 0006} (2000) 009
[arXiv:hep-ph/9909540].

\bibitem{deltamt}
S.~Heinemeyer, S.~Kraml, W.~Porod and G.~Weiglein,
arXiv:hep-ph/0306181.

\bibitem{bsgth}
G.~Degrassi, P.~Gambino and G.~F.~Giudice,
JHEP {\bf 0012} (2000) 009
[arXiv:hep-ph/0009337];\\
M.~Carena, D.~Garcia, U.~Nierste and C.~E.~Wagner,
Phys.\ Lett.\ B {\bf 499} (2001) 141
[arXiv:hep-ph/0010003];\\
D.~A.~Demir and K.~A.~Olive,
Phys.\ Rev.\ D {\bf 65} (2002) 034007
[arXiv:hep-ph/0107329].

\bibitem{Degrassi:1998es}
G.~Degrassi and G.~F.~Giudice,
Phys.\ Rev.\ D {\bf 58} (1998) 053007
[arXiv:hep-ph/9803384].

\bibitem{gm2} 
H.~N.~Brown {\it et al.}  [Muon g-2 Collaboration],
Phys.\ Rev.\ Lett.\  {\bf 86} (2001) 2227
[arXiv:hep-ex/0102017];\\
G.~W.~Bennett {\it et al.}  [Muon g-2 Collaboration],
Phys.\ Rev.\ Lett.\  {\bf 89} (2002) 101804
[Erratum-ibid.\  {\bf 89} (2002) 129903]
[arXiv:hep-ex/0208001].

\bibitem{gm2disc} F.~Jegerlehner, these proceedings.

\bibitem{cdm} 
A.~Melchiorri and J.~Silk,
Phys.\ Rev.\ D {\bf 66} (2002) 041301
[arXiv:astro-ph/0203200];\\
A.~Benoit {\it et al.}  [the Archeops Collaboration],
Astron.\ Astrophys.\  {\bf 399} (2003) L25
[arXiv:astro-ph/0210306];\\
J.~R.~Ellis, J.~S.~Hagelin, D.~V.~Nanopoulos, K.~A.~Olive and M.~Srednicki,
Nucl.\ Phys.\ B {\bf 238} (1984) 453;\\
H.~Goldberg,
Phys.\ Rev.\ Lett.\  {\bf 50} (1983) 1419.


\bibitem{bsg} S.~Ahmed et al., {CLEO CONF 99-10};\\
  $[$Belle Collaboration$]$, BELLE-CONF-0003, contribution to the 30th 
  International conference on High-Energy Physics, Osaka, 2000.
  See also\\
K.~Abe {\it et al.}  [Belle Collaboration],
Phys.\ Lett.\ B {\bf 511} (2001) 151
[arXiv:hep-ex/0103042];\\
K.~Abe {\it et al.}  [Belle Collaboration],
arXiv:hep-ex/0107065;\\
L.~Lista  [BABAR Collaboration],
arXiv:hep-ex/0110010.

\bibitem{ellisnew}
J.~R.~Ellis, K.~A.~Olive, Y.~Santoso and V.~C.~Spanos,
arXiv:hep-ph/0303043.

\bibitem{ehow2} 
J.~R.~Ellis, S.~Heinemeyer, K.~A.~Olive and G.~Weiglein,
JHEP {\bf 0301} (2003) 006
[arXiv:hep-ph/0211206].

\bibitem{gm2mSUGRA} 
W.~de Boer, M.~Huber, C.~Sander and D.~I.~Kazakov,
Phys.\ Lett.\ B {\bf 515} (2001) 283.

\bibitem{MWMTplot}
S.~Heinemeyer and G.~Weiglein,
arXiv:hep-ph/0012364.

\bibitem{gigaz}
J.~Erler, S.~Heinemeyer, W.~Hollik, G.~Weiglein and P.~M.~Zerwas,
Phys.\ Lett.\ B {\bf 486} (2000) 125
[arXiv:hep-ph/0005024].

\bibitem{alr}
M.~Swartz, these proceedings.

\bibitem{afb}
P.~Wells, these proceedings.

\bibitem{chanowitz} 
M.~S.~Chanowitz,
arXiv:hep-ph/0304199, these proceedings.

\bibitem{lightsf}
G.~Altarelli, F.~Caravaglios, G.~F.~Giudice, P.~Gambino and G.~Ridolfi,
JHEP {\bf 0106} (2001) 018
[arXiv:hep-ph/0106029].

\bibitem{mhLEPfinal} 
G.~Abbiendi et al.\  [ALEPH, DELPHI, L3, OPAL Collaborations],
arXiv:hep-ex/0306033, to appear in {\em Phys. Lett.} {\bf B}.

\bibitem{nutev}
K.~McFarland, these proceedings.

\bibitem{Davidson:2001ji}
S.~Davidson, S.~Forte, P.~Gambino, N.~Rius and A.~Strumia,
JHEP {\bf 0202} (2002) 037
[arXiv:hep-ph/0112302].

\bibitem{ewpoMSSM} 
W.~de Boer and C.~Sander,
appeared in the proceedings of ``SUSY 02'', DESY Hamburg, Germany,
June 17-23, 2002, p.\ 1121-1126.

\end{thebibliography}
